\documentclass{article}
\PassOptionsToPackage{numbers, compress}{natbib}

\usepackage[final]{neurips_2023}

\usepackage[utf8]{inputenc} 
\usepackage[T1]{fontenc}    
\usepackage{hyperref}       
\usepackage{url}            
\usepackage{booktabs}       
\usepackage{amsfonts}       
\usepackage{nicefrac}       
\usepackage{microtype}      
\usepackage{xcolor}         
\usepackage{graphicx}

\title{ProsDectNet: Bridging the Gap in Prostate Cancer Detection via Transrectal B-mode Ultrasound Imaging}
\author{%
  \textbf{Sulaiman Vesal}\textsuperscript{1,2}, \textbf{Indrani Bhattacharya}\textsuperscript{2}, \textbf{Hassan Jahanandish}\textsuperscript{1}, \textbf{Xinran Li}\textsuperscript{3}, \\ \textbf{Zachary Kornberg}\textsuperscript{1}, \textbf{Steve Ran Zhou}\textsuperscript{1}, \textbf{Elijah Richard Sommer}\textsuperscript{4}, \textbf{Moon Hyung Choi}\textsuperscript{1}, \\ \textbf{Richard E. Fan}\textsuperscript{1}, \textbf{Geoffrey A. Sonn}\textsuperscript{1,*}, and \textbf{Mirabela Rusu}\textsuperscript{1,2}\thanks{Equal contribution as senior authors.} \\ \\
 \textsuperscript{1}Department of Urology, Stanford University, Stanford, CA, 94305, USA \\
 \textsuperscript{2} Department of Radiology, Stanford University, Stanford, CA, 94305, USA \\
 \textsuperscript{3} Institute for Computational and Mathematical Engineering, Stanford University, \\ Stanford, CA, 94305, USA \\
 \textsuperscript{4} School of Medicine, Stanford University, Stanford, CA, 94305, USA \\ 
  \texttt{svesal@stanford.edu} \\
}
\begin{document}

\maketitle
\begin{abstract}
Interpreting traditional B-mode ultrasound images can be challenging due to image artifacts (e.g., shadowing, speckle), leading to low sensitivity and limited diagnostic accuracy. While Magnetic Resonance Imaging (MRI) has been proposed as a solution, it is expensive and not widely available. Furthermore, most biopsies are guided by Transrectal Ultrasound (TRUS) alone and can miss up to 52\% cancers, highlighting the need for improved targeting. To address this issue, we propose ProsDectNet, a multi-task deep learning approach that localizes prostate cancer on B-mode ultrasound. Our model is pre-trained using radiologist-labeled data and fine-tuned using biopsy-confirmed labels. ProsDectNet includes a lesion detection and patch classification head, with uncertainty minimization using entropy to improve model performance and reduce false positive predictions. We trained and validated ProsDectNet using a cohort of 289 patients who underwent MRI-TRUS fusion targeted biopsy. We then tested our approach on a group of 41 patients and found that ProsDectNet outperformed the average expert clinician in detecting prostate cancer on B-mode ultrasound images, achieving a patient-level ROC-AUC of 82\%, a sensitivity of 74\%, and a specificity of 67\%. Our results demonstrate that ProsDectNet has the potential to be used as a computer-aided diagnosis system to improve targeted biopsy and treatment planning. 
\end{abstract}

\section{Introduction}
Transrectal ultrasound-guided (TRUS) biopsy procedures are commonly used for diagnosing prostate cancer \cite{Sonn2013}. While B-mode TRUS images enable urologists to guide the biopsy needle in real-time, their low sensitivity in detecting prostate cancer due to shadowing artifacts and low signal-to-noise ratio remains a challenge \cite{Gaffney2021,Azizi2018b}. Although MRI has improved the accuracy of prostate biopsies through fusion with TRUS, MRI remains underutilized due to limited accessibility and lack of interpretation expertise. Thereby, ultrasound remains the most common imaging modality for prostate cancer diagnosis \cite{AhmedHashimU2017Daom}, being used in 93\% of biopsy procedures in the absence of MRI \cite{Gaffney2021}. However, such TRUS-only biopsy procedures miss up to 52\% of clinically significant cancers \cite{AhmedHashimU2017Daom} due to systematic sampling of the prostate without targeting suspicious regions. Therefore, it is crucial to improve targeting during the biopsy procedure by localizing suspicious regions on the ubiquitous B-mode ultrasound image and enabling early detection of prostate cancer. 

The interpretation of prostate B-mode ultrasound images is challenging, only 70\% of cancers are hypoechoic, and many confounders look like cancer (hypoechoic) but correspond to normal tissue. These limitations have been addressed clinically by introducing more advanced ultrasound-based modalities and using machine learning for analysis \cite{Azizi2018b,Imani2015,Azizi2018,Azizi2018c,Akatsuka2022,Sedghi2019}. However, these modalities are not routinely available and are considered investigational. Few studies have used the B-mode ultrasound images, but trained machine learning methods in small biased populations where all patients had clinically significant cancer \cite{Han2008,Wildeboer2020}. Training deep learning methods to localize cancer on B-mode ultrasound images has two challenges. First, generating accurate cancer labels on ultrasound images is tedious and suffers from inter-reader variability. Second, the many confounders of cancer cause models to have significant false positives. To overcome these challenges, we propose ProsDectNet, a prostate lesion detection framework. The model is pre-trained using weak labels from a public dataset \cite{Natarajan2020}, which only includes radiologist labels, and then fine-tuned using pathology-confirmed biopsy labels. 

\section{Methods and Materials}
\textbf{Data description:} Our study, approved by the institutional review board, focused on patients who underwent MRI-TRUS fusion targeted biopsy utilizing the Artemis system. We organized our subjects into two cohorts for analysis. Cohort C1 comprised 330 patients from our institution who underwent biopsy procedures. Among them, 123 patients also underwent radical prostatectomy, while 207 patients had negative biopsy results. Cohort C2 included 1,151 patients sourced from the Prostate-MRI-US-Biopsy cohort, accessible in the Cancer Imaging Archive~\cite{Natarajan2020}. This dataset contained weak-label annotations by radiologists, crucial for pretraining ProsDectNet. For our study, we conducted training and validation using 252 and 37 patients from our internal cohort, respectively. To assess the performance of our models, we employed a test set consisting of 41 patients. Quantitative assessments of the models were conducted at both lesion-level and patient-level similar to \cite{globalstat}, employing various metrics such as the Receiver Operating Characteristic curve (ROC-AUC), sensitivity (SE), specificity (SP), negative predictive value (NPV), positive predictive value (PPV), and accuracy (ACC) for comprehensive evaluation.

\begin{figure}[!t]
    \centering
    \includegraphics[width=0.8\textwidth]{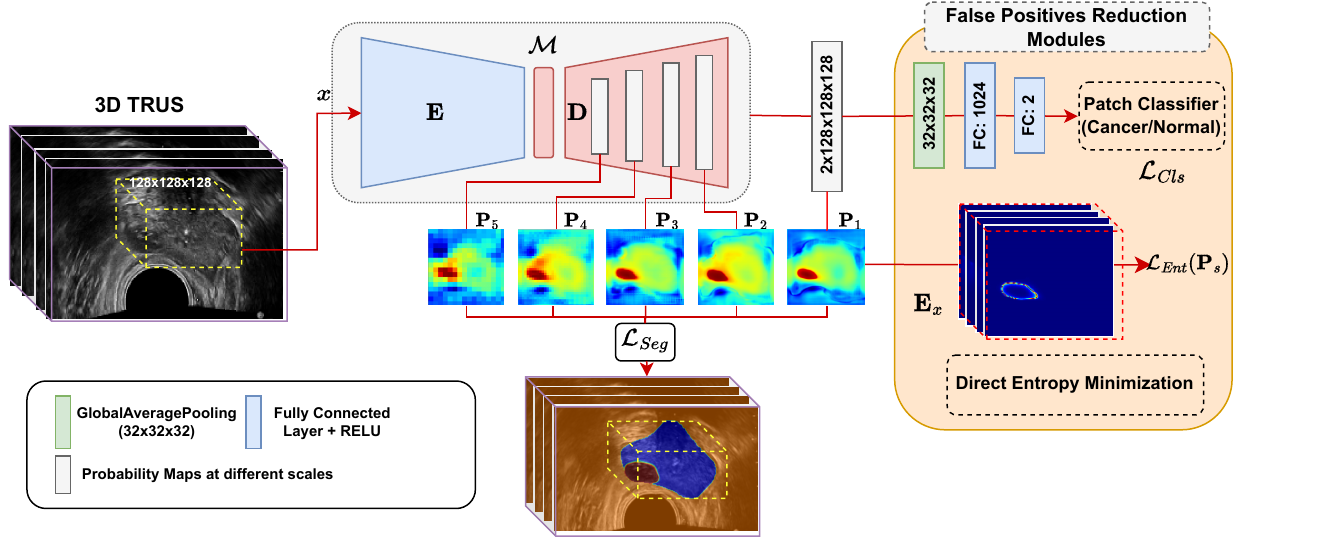}
    \caption{ProsDectNet network archticture.}
    \label{fig:flowchart}
\end{figure}

\textbf{ProsDectNet} Fig. \ref{fig:flowchart} illustrates the architecture of ProsDectNet, designed for localizing and detecting prostate cancer. Initially, ProsDectNet undergoes a pre-training phase utilizing weak labels from Cohort C2. Following this, it is fine-tuned using strong labels from Cohort C1. During training, ProsDectNet randomly extracts 3D patches ($128\times128\times128$) from the prostate region to train the detection model denoted as $\mathcal{M}$. The model $\mathcal{M}$ consists of an encoder with pyramid predictions at multiple scales, enabling deep supervision. To enhance precision and minimize false positive predictions, two crucial components are integrated: a classification head that determines the presence or absence of cancer in a patch, and an uncertainty estimation head that computes the entropy map $\mathbf{E}_x$ from $\mathbf{P}_x$. The entropy map provides information about prediction uncertainty. For the backbone model of ProsDectNet, we employed 3D-UNet \cite{3DUNet2016} with additional auxiliary detection heads at various resolution levels of the decoder \cite{Lin_2017_CVPR}. This design choice enables ProsDectNet to leverage multi-scale information effectively. During training, for a patch $x$, the model $\mathcal{M}(x)$ produces a set of multi-scale predictions $[\mathbf{p'}_{1}, \mathbf{p'}_{2}, ..., \mathbf{p'}_{s}]$, where $\mathbf{p'}_{s}$ represents the prediction at the $s$-th scale. Here, a smaller $s$ indicates a higher resolution. We denote these rescaled multi-scale predictions as $[\mathbf{P}{1}, \mathbf{P}_{2}, ..., \mathbf{P}_{s}]$. To address the significant class imbalance, where cancer regions constitute only 4\% of all prostate voxels, $\mathcal{M}$ is trained using $\mathcal{L}_{seg}(\mathbf{P}_{s}, y)$, a weighted combination of Dice coefficient loss and focal loss \cite{YEUNG2022102026} at different scales, effectively mitigating this imbalance.

\textbf{False Positive Reduction.} Since the ultrasound images include many hypoechoic regions (some that are cancer and some that are not), we introduced two modules aimed at reducing false positives while preserving the model's sensitivity. \textbf{Patch Classification.} We added a classification head to $\mathcal{M}$ to create a multi-task model. The primary objective was to train $\mathcal{M}$ not only to segment the patches but also to classify them as cancerous or normal, thereby providing additional supervision to mitigate false positive predictions. To accomplish this, we applied a global average pooling layer with a kernel size of $32\times32\times32$ to the probability output of the model's prediction $\mathbf{P}_{x}$ and added two fully connected layers to estimate the probability of cancer for each patch. The classification head used a softmax function and trained using cross-entropy loss denoted as $\mathcal{L}_{Cls}$.
\textbf{Entropy Minimisation.} To address the issue of uncertain predictions and reduce false positives, we have adopted an entropy-driven approach similar to that proposed in~\cite{Vu2019}. By minimizing the entropy of the predicted probabilities, we encourage $\mathcal{M}$ to make more confident and accurate predictions. Given the voxel-wise probability prediction $\mathbf{P}_{s}$ of input image ${x}$, we use Shannon Entropy~\cite{Vu2019} to calculate the entropy map at the voxel-level: $\mathcal{L}_{Ent}({\mathbf{P}_{s}})=-\frac{1}{N}\sum_{n=1}^{N}\sum_{c=1}^{C}\mathbf{P}_{s}^{n,c}\log \mathbf{P}_{s}^{n,c}$. Here, $C$ denotes the number of classes (e.g., background and lesion), $N$ denotes the number of images, and $\mathbf{P}_{s}^{n,c}$ is the predicted probability of the voxel belonging to class $c$. 

\textbf{{Overall Objective Function.}} The ProsDectNet trained by minimizing the following combined objective function:
$
    \mathcal{L}_{total} = \mathcal{L}_{seg} + \lambda_{1}\mathcal{L}_{Cls} + \lambda_{2}\mathcal{L}_{Ent}
$
where $\mathcal{L}_{seg}$ is a joint dice loss and focal loss. $\mathcal{L}_{Cls}$ is the patch classification loss and $\mathcal{L}_{Ent}$ for entropy loss. $\lambda_{1}$ and $\lambda_{2}$ are the weights for each loss term, which were set to 1.0 and 0.2 in our experiments. 

\section{Results}

Previous studies on prostate cancer detection using B-mode ultrasound were limited due to non-open-sourced methods and internal cohorts, hindering direct comparisons with our approach. To assess ProsDectNet's performance, we compared it with state-of-the-art segmentation models: 3D-UNet \cite{3DUNet2016}, UNETR \cite{Hatamizadeh2021}, and SwinUNTER \cite{Hatamizadeh2022}. We ensured fairness by applying identical pre and post-processing techniques to all models. Table \ref{tab:annotator} presents the evaluation results, encompassing both lesion-level and patient-level assessments. ProsDectNet, utilizing pre-trained weights, a classification head, and entropy loss, exhibited superior performance. The model achieved a lesion-level sensitivity of 66.0\%, specificity of 90.0\%, PPV of 74.0\%, and accuracy of 85.0\% (P-value $=$ 0.001). Notably, ProsDectNet outperformed the average expert, achieving a sensitivity of 74.0\% and NPV of 74.0\% on the patient-level, surpassing the performance of most individual experts. Intriguingly, significant intra-annotation variability was observed among clinicians. Expert 3, a radiologist with 12 years of experience, exhibited the highest specificity but lower sensitivity than ProsDectNet. These findings underline ProsDectNet's potential for accurate prostate cancer detection in TRUS images, showcasing its robustness against inter-observer variability.

\begin{figure}[!t]
    \centering
    \includegraphics[width=0.95\textwidth]{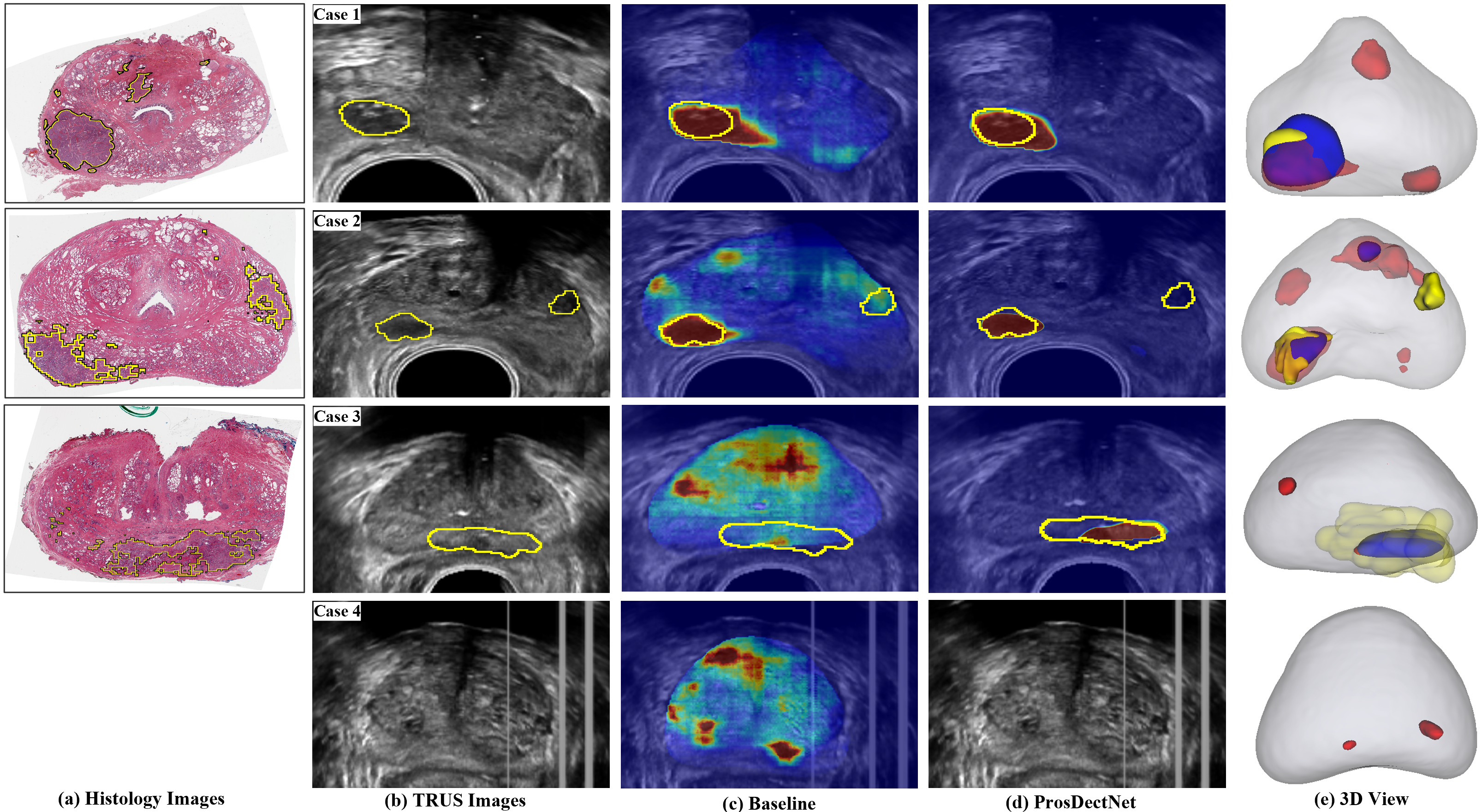}
    \caption{Prediction probability maps for four patients from the test set. Ground truth labels are shown in yellow color. The first column shows the corresponding histology slices. The last column shows the 3D visualization of ProsDectNet prediction (blue) and the baseline model (red).}
    \label{fig:my_label}
\end{figure}

\textbf{Qualitative Evaluation.} Fig. \ref{fig:my_label} shows the entropy map prediction outputs of the baseline model (3D-UNet) and ProsDectNet for four patients from the test set, along with their ground-truth labels (yellow color). Additionally, the 3D visualization of the predictions is presented for qualitative interpretation. ProsDectNet outperformed the baseline model (3D-UNet) in detecting cancer lesions with fewer false positive predictions and yielding the lowest uncertainty. The baseline model wrongly identified shadowing artifacts (cases 3 \& 4) in the transition zone as cancer due to their similar appearance to hypo-echoic cancer regions and missed the lesion (case 3). 

\begin{table}[!t]
    \centering
    \caption{ Lesion-level and patient-level evaluations between experts, other models and ProsDectNet.}
    \label{tab:annotator}
    \resizebox{0.72\textwidth}{!}{
    \begin{tabular}{lcccccc|ccccc}
    \hline
    & \multicolumn{6}{c}{\textbf{Lesion-Level}} & \multicolumn{5}{|c}{\textbf{Patient-Level}} \\ \cline{2-12}
    \textbf{Annotator} & \textbf{ROC-AUC} & \textbf{SE} & \textbf{SP} & \textbf{PPV} & \textbf{NPV}  & \textbf{ACC} & \textbf{SE} & \textbf{SP} & \textbf{PPV} & \textbf{NPV}  & \textbf{ACC}\\
    \hline
     Expert 1  & 0.64 & 0.41 & 0.91 & 0.63 & 0.89 & 0.84 & 0.47 & 0.86  & 0.75 & 0.64 & 0.68\\
     Expert 2  & 0.74 & 0.54 & 0.86 & 0.82 & 0.91 & 0.80 & 0.63 & 0.38 & 0.48 & 0.53 & 0.50\\
     Expert 3  & 0.73 & 0.54 & \textbf{0.94} & 0.\textbf{96} & 0.90 & \textbf{0.89} & 0.63 & \textbf{0.90} & \textbf{0.86} & 0.73 & \textbf{0.78}\\
     Expert 4  & 0.60 & 0.23 & 0.92 & 0.75 & 0.85 & 0.80 & 0.26 & 0.67 & 0.40 & 0.48 & 0.48\\
     \hline
     Average Expert:  & 0.68 & 0.43 & 0.91 &  0.79 & 0.89 & 0.83 & 0.50 & 0.70 & 0.62 & 0.60 & 0.61\\
    \hline
    3D-UNet \cite{3DUNet2016} & 0.71  & 0.52 & 0.72& 0.41 & 0.90  & 0.72  & 0.58 & 0.71 & 0.65  & 0.65 & 0.65\\
    UNETR \cite{Hatamizadeh2021} & 0.56  & 0.39 & 0.62 & 0.27 & 0.80  & 0.58  & 0.50 & 0.55 & 0.52 & 0.52 & 0.53\\
    SwinUNETR \cite{Hatamizadeh2022} & 0.69  & 0.69 & 0.62 & 0.40 & 0.91  & 0.61  & 0.74 & 0.43 & 0.54 & 0.64 & 0.58\\
    ProsDectNet (ours)  & \textbf{0.82} & \textbf{0.66} & 0.90 & 0.74 & \textbf{0.93} & 0.85 & \textbf{0.74} & 0.67  & 0.67 & \textbf{0.74} & 0.70\\
    \hline
    \end{tabular}}
\end{table}

\section{Conclusion}
This paper presents ProsDectNet, a novel deep learning framework for detecting prostate cancer on B-mode TRUS images. Our proposed approach involves multi-task lesion detection and classification using deep supervision and leverages entropy to reduce uncertainty and false positive predictions, thereby improving the overall performance and addressing the challenges of ultrasound images. To assess our approach, we compared the performance of ProsDectNet against other architectures as well as four clinical experts with varying levels of experience and found it outperforms them in terms of sensitivity and specificity when compared to the average expert. These results demonstrate the potential of ProsDectNet as a tool to aid in clinical diagnosis and biopsy targeting using B-mode ultrasound alone, particularly in situations where MRI may not be available. Future work will involve expanding the dataset and evaluating the system in clinical settings, to further validate the efficacy of our approach.

\section{Acknowledgements}
We acknowledge the following funding sources: Departments of Radiology and Urology, Stanford University and  International Alliance for Cancer Early Detection (ACED). The research reported in this publication was supported by the National Cancer Institute of the National Institutes of Health under Award Number R37CA260346. The content is solely the responsibility of the authors and does not necessarily represent the official views of the National Institutes of Health.

\bibliographystyle{plainnat}
\bibliography{biblo.bib}
%

\section*{Supplementary Materials}

\begin{table}[ht]
    \centering
    \caption{Ablation study for lesion-level and patient-level evaluations on test set and the impact of each loss componenet on overall performance. AUC-ROC: Area under curve, SE: sensitivity, SP: Specificity, PPV: Positive Predictive Value, NPV: Negative predicted value, ACC: Accuracy.}
    \label{tab:lesion_level_eval}
    \resizebox{\textwidth}{!}{
    \begin{tabular}{lcccccc|ccccccc}
    \hline
       & \multicolumn{6}{c}{\textbf{Lesion-Level}} & \multicolumn{5}{|c}{\textbf{Patient-Level}} \\ \cline{2-12}
    \textbf{Model} & \textbf{ROC-AUC} & \textbf{SE} & \textbf{SP} & \textbf{PPV} & \textbf{NPV}  & \textbf{ACC} &  \textbf{SE} & \textbf{SP} & \textbf{PPV} & \textbf{NPV} & \textbf{ACC} \\
    \hline
    \textbf{ProsDectNet} & 0.76  & 0.67 & 0.60 & 0.48 & 0.90  & 0.61  & 0.73 & 0.23  & 0.46 & 0.50 & 0.47\\
    + Pretrained + $\mathcal{L}_{Cls}$ & 0.70  & \textbf{0.70} & 0.78 & 0.62 & 0.93 & 0.77  & \textbf{0.79 } & 0.43  & 0.56 & 0.69 & 0.60\\
    + Pretrained + $\mathcal{L}_{Ent}$ & 0.76  & 0.62 & 0.77 & 0.63 & 0.89 & 0.76  & 0.74 & 0.57  & 0.61 & 0.71 & 0.65\\
    + Pretrained + $\mathcal{L}_{Cls}$ + $\mathcal{L}_{Ent}$ & \textbf{0.82}  & 0.66 & \textbf{0.90} & \textbf{0.74} & \textbf{0.93} &  \textbf{0.85}  & 0.74 & \textbf{0.67}  & \textbf{0.67} & \textbf{0.74} & \textbf{0.70}\\
    \hline
    \end{tabular}}
\end{table}

\begin{figure}[ht]
    \centering
    \includegraphics[scale=0.78]{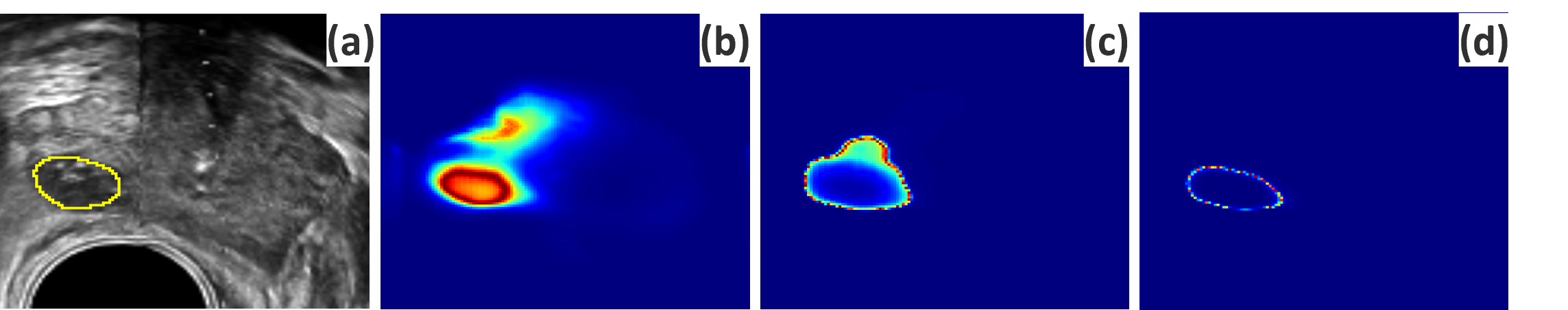}
    \caption{Entropy maps for different models. (a) TRUS slice, (b) ProsDectNet with classification head, (c) with entropy loss, and (d) with classification head and entropy loss. }
    \label{fig:ent}
\end{figure}




\end{document}